\documentclass[preprint,journal]{vgtc}            

\onlineid{1335}

\vgtccategory{Research}

\title{\tool: Interactive Visual Diagnosis of Behavioral Inconsistencies in LLM-based Agentic Systems}

\author{
  Shuo Yan, Xiaolin Wen, Shaolun Ruan, Yanjie Zhang, Jiaming Mi, Yushi Sun, Huamin Qu, and Rui Sheng*
}

\authorfooter{
  \item 
    Shuo Yan and Jiaming Mi are with East China Normal University. 
    E-mail: 10245102482@stu,ecnu.edu.cn and jmmi@stu.ecnu.edu.cn.
  \item
  	Xiaolin Wen is with Nanyang Technological University.
  	E-mail: xiaolin004@e.ntu.edu.sg.
  \item 
    Shaolun Ruan is with Singapore Management University. 
    E-mail: haywardryan@foxmail.com.
  \item Yanjie Zhang, Yushi Sun, Huamin Qu, and Rui Sheng are with the Hong Kong University of Science
and Technology. 
    E-mail: {yzhangvj,ysunbp,huamin, rshengac}@connect.ust.hk.
  \item *: Corresponding Author
}

\usepackage{xspace}
\usepackage{graphicx}

\newcommand{\tool}{\textit{InconLens}\xspace}

\abstract{%
 Large Language Model (LLM)-based agentic systems have shown growing promise in tackling complex, multi-step tasks through autonomous planning, reasoning, and interaction with external environments. However, the stochastic nature of LLM generation introduces intrinsic behavioral inconsistency: the same agent may succeed in one execution but fail in another under identical inputs. Diagnosing such inconsistencies remains a major challenge for developers, as agent execution logs are often lengthy, unstructured, and difficult to compare across runs. Existing debugging and evaluation tools primarily focus on inspecting single executions, offering limited support for understanding how and why agent behaviors diverge across repeated runs.
 To address this challenge, we introduce \tool, a visual analytics system designed to support interactive diagnosis of LLM-based agentic systems with a particular focus on cross-run behavioral analysis. \tool introduces information nodes as an intermediate abstraction that captures canonical informational milestones shared across executions, enabling semantic alignment and inspection of agent reasoning trajectories across multiple runs. We demonstrate the effectiveness of \tool through a detailed case study and further validate its usability and analytical value via expert interviews. Our results show that \tool enables developers to more efficiently identify divergence points, uncover latent failure modes, and gain actionable insights into improving the reliability and stability of agentic systems.
}


\graphicspath{{figs/}{figures/}{pictures/}{images/}{./}} 

\usepackage{tabu}                      
\usepackage{booktabs}                  
\usepackage{lipsum}                    
\usepackage{mwe}                       
\usepackage{ccicons}                   

\usepackage{mathptmx}                  

\begin{document}



\maketitle

\section{Introduction}

There is a rapidly growing interest in organizing Large Language Models (LLMs) into agentic architectures to address complex, multi-step tasks that go beyond single-turn prompting~\cite{park2023generative, xi2023rise, shen2026storylensedu, lin2025survey}. In these architectures, LLMs are framed as autonomous or semi-autonomous agents that can plan goals, decompose tasks, reason over intermediate states, and take actions through iterative interactions with external environments, tools, or other agents. This paradigm has already demonstrated promising potential across a wide range of domains, such as web searching~\cite{deng2023mind2web, zhou2024webarena}, scientific discovery~\cite{boiko2023autonomous, bran2023chemcrow, romera2023mathematical, shi2026survey}, and software development~\cite{qian2023chatdev, hong2024metagpt}. Together, these advances suggest that LLM-based agentic systems are emerging as a powerful computational paradigm for tackling complex, real-world problems.

Despite the growing capabilities of LLM-based agentic systems, developing robust and reliable agentic applications remains a significant challenge. LLM-based agents inherently operate under stochastic, probabilistic generation processes, which introduce intrinsic uncertainty into their behaviors. As a result, agents may succeed in completing 
a task in one execution but fails in subsequent runs.
Such behavioral inconsistencies compel developers to carefully inspect and compare agent behaviors across multiple runs. 
By carefully examining agent plans and execution details, developers aim to identify where and how reasoning trajectories diverge, uncover latent failure modes, and derive actionable insights for systematically improving agentic systems.

However, diagnosing behavioral inconsistencies in LLM-based agentic systems remains challenging due to two intertwined factors. First, semantic alignment across heterogeneous execution trajectories is difficult.
Specifically, different runs may involve entirely different sequences of agents, tools, and strategies to achieve the same sub-goal. Therefore, meaningful comparison requires abstracting away incidental execution details while preserving task-relevant milestones across different runs.
Second, identifying behavior patterns associated with success or failure requires fluid movement between overview and detail across multiple runs. Developers must locate divergent task stages and understand their intents, inspect the underlying agent actions, and validate hypotheses against raw execution evidence. While automated logs or LLM-generated summaries can aid single-run analysis, they are insufficient for cross-run comparative reasoning, which demands simultaneous inspection of multiple aligned trajectories and is better supported by interactive visual analytics approaches.

To address this problem, we conduct a formative study with five experienced developers. 
Based on the insights derived from the semi-structured interviews, we propose \tool, a visual analytics system designed to support the diagnosis of LLM-based agentic systems.
This system focuses on analyzing and understanding behavioral inconsistencies in centralized multi-agent systems, which is a popular agentic paradigm where a central orchestration mechanism coordinates multiple agents to collaboratively perform complex tasks~\cite{fourney2024magentic,chen2024agentverse,camel2023li,shen2023hugginggpt}.
Specifically, we introduce \textit{information nodes} as conceptual intermediaries between developers and raw agent logs.
Information nodes capture canonical informational milestones abstracted across multiple runs.
They enable developers to align, compare, and inspect agent behaviors across executions, significantly reducing the cognitive burden imposed by verbose and unstructured logs. We demonstrate the effectiveness of \tool through a detailed case study and further validate its usability and analytical value via expert interviews.
Overall, our contributions can be summarized as follows:
\begin{itemize}[leftmargin=*]
    \item We conduct a formative study with five experienced AI developers to identify key design requirements for diagnosing behavioral inconsistencies in LLM-based agentic systems  
    \item We propose a visual analytics system, \tool, which integrates \textit{information nodes} to better support cross-run alignment and diagnosis of inconsistent agent behaviors in centralized multi-agent systems.
    \item We validate the effectiveness of \tool through a detailed case study and expert interviews, and further discuss implications and future research directions.
\end{itemize}
\section{Background}
In this section, we first introduce the background of LLM-based agentic systems. After that, we elaborate on the current diagnostic tools for such systems, clarifying their limitations in cross-run comparison for inconsistency identification.

\subsection{LLM-Based Agentic Systems}
The integration of Large Language Models (LLMs) into autonomous frameworks has catalyzed the emergence of agentic AI. This marks a fundamental transition from static, prompt-driven generation to dynamic, goal-directed problem solving~\cite{xi2023rise, Acharya2025AgenticAI, Bandi2025TheRO}. Unlike traditional software systems governed by deterministic rules or simple generative models, agentic systems utilize LLMs as cognitive controllers (``brains'') to perceive complex environments, reason about abstract objectives, and orchestrate actions through external tools~\cite{wang2024agentailanggraphmodular, abouali2025agentic, arunkumar2026agentic}. This paradigm shift enables systems to operate with varying degrees of autonomy, ranging from human-in-the-loop copilots to fully independent agents capable of long-horizon planning, self-correction, and dual-paradigm (neuro-symbolic) reasoning~\cite{Hughes2025AIAgents, Plaat2025AgenticLLM, sapkota2026agentic}.

\begin{figure}[h]
  \centering
  \includegraphics[width=\columnwidth]{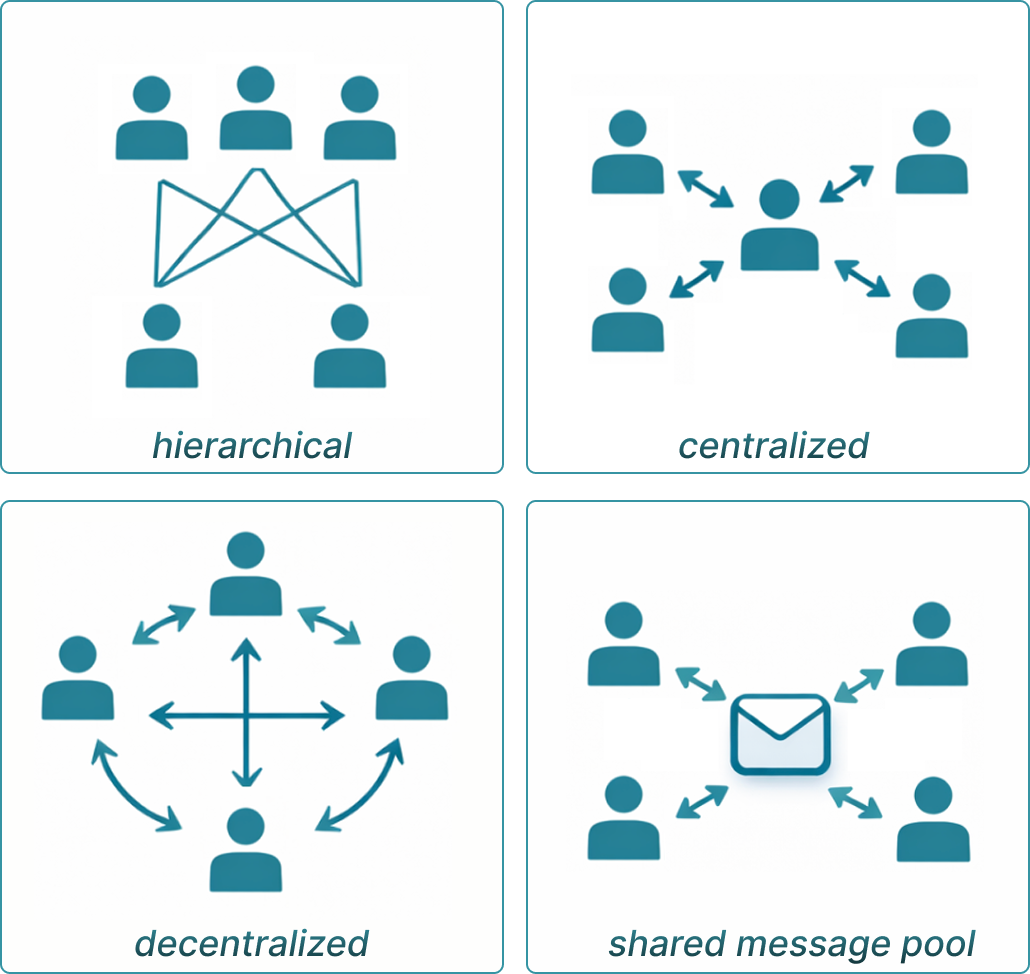}
  \caption{Four interaction architectures for LLM-based multi-agent systems.}
  \label{fig:arche}
\end{figure}

Based on the architectural composition and collaborative scope, recent agentic systems are divided into two categories: single-agent systems and multi-agent systems. 
Single-agent systems rely on a single LLM-based agent that autonomously performs task planning, reasoning, and tool use through iterative interactions with the environment~\cite{yao2023react, wang2023voyager}.
While such systems can handle moderately complex workflows, their capabilities are often limited by the lack of specialization and the bounded reasoning capacity of a single model.
By contrast, multi-agent systems leverage a set of specialized agents with distinct roles, enabling them to distribute workloads and combine complementary capabilities, often outperforming single-agent counterparts in handling complex queries~\cite{chen2024agentverse, li2025newera, shen2024fromdata, zhang2024seewidely, sapkota2026agentic, Bandi2025TheRO, qian2023chatdev, hong2024metagpt}. 
Additionally, within multi-agent systems, interactions are further structured into diverse topologies to optimize coordination and information flow, including centralized, decentralized, hierarchical, and shared message pool~\cite{ aryal2026survey, Yao2025Survey} (\autoref{fig:arche}).
In a hierarchical architecture, control is distributed across multiple levels, with higher-level agents dynamically coordinating lower-level specialists, as demonstrated by MDAgents~\cite{kim2024mdagents}.
A decentralized architecture removes global control altogether, allowing agents to negotiate plans through direct, peer-to-peer interactions, as in DMAS~\cite{chen2024scalable}.
By contrast, a centralized architecture concentrates coordination in a single agent that collects intermediate results and directs all other agents, a design adopted by Magentic-One~\cite{fourney2024magentic}.
Finally, a shared message pool architecture enables coordination through a common communication space that agents can asynchronously read from and write to, as exemplified by MetaGPT~\cite{hong2024metagpt}.
In this paper, we focus on diagnosing failures in centralized multi-agent systems.

\subsection{Automated Diagnosis of Agentic Systems}
The rapid evolution of AI agents from simple chatbots to autonomous systems capable of long-horizon, multi-step tasks has introduced substantial challenges in debugging and reliability. To address these issues, a growing body of diagnosis algorithms has been proposed for both single-agent and multi-agent systems, spanning trajectory-based analysis~\cite{barke2026agentrxdiagnosingaiagent,huang2026tracecoder}, multi-agent collaboration frameworks~\cite{lee2025unidebugger,krishnamoorthy-etal-2025-multi}, and industrial-scale evaluation~\cite{ahmed2026specops}. Representative systems include AgentRx~\cite{barke2026agentrxdiagnosingaiagent}, which provides a structured, domain-agnostic framework to pinpoint critical failure steps. TraceCoder~\cite{huang2026tracecoder} also introduces a trace-driven multi-agent approach for debugging LLM-generated code. UniDebugger~\cite{lee2025unidebugger} proposes an end-to-end unified debugging framework based on hierarchical multi-agent synergy. Additionally, SpecOps~\cite{ahmed2026specops} presents a fully automated evaluation approach for GUI-based, product-level LLM agents operating in real-world environments rather than simulators. However, despite these advances, the limited performance of automated diagnosis, particularly under constraints of deep inspection and contextual understanding, means that developers remain indispensable for actively detecting failures and interpreting their root causes.

\subsection{Interactive Agentic System Diagnosis Tools}
A complementary class of tools emphasizes human-in-the-loop interaction to enable developers to inspect, manipulate, and collaboratively debug agent behaviors~\cite{WaitGPT, lu2025AgentLens, hutter2026agentstepperinteractivedebuggingsoftware, 10.1145/3706598.3713581, sheng2026dills, ma2026retrace}.
Those works have begun to address the unique challenges of debugging LLM-based agentic systems by shifting from low-level log analysis to high-level interactive steering.
For example, AgentStepper~\cite{hutter2026agentstepperinteractivedebuggingsoftware} focuses on debugging software development agents by treating agent trajectories as structured conversations, allowing developers to set breakpoints and perform stepwise execution to identify bugs in the agent's orchestrating program. AGDebugger~\cite{10.1145/3706598.3713581} targets multi-agent collaboration, providing interactive message resets and a branching graph visualization to help developers localize critical failure steps and steer agent behavior during complex, stateful interactions. Furthermore, DiLLS~\cite{sheng2026dills} proposes a hierarchical approach to debug multi-agent behaviors, organizing complex logs into ``activities, actions, and operations'' to reduce cognitive load during root-cause diagnosis. ReTrace~\cite{ma2026retrace} specifically debugs the internal reasoning traces of Large Reasoning Models (LRMs) to uncover patterns like overthinking or early task abandonment.
However, while existing interactive diagnosis systems are effective at detecting logic errors, tool misuse, or planning failures within individual execution traces, they are limited in addressing the cross-run diagnostic challenge: understanding how and why agent behaviors vary across repeated executions of the same task.
\section{Design Requirements}
Our goal in designing \tool is to help users diagnose inconsistent behaviors across repeated runs of LLM-based multi-agent systems, with a particular emphasis on understanding how inter-agent coordination patterns in centralized multi-agent systems contribute to divergent outcomes. In practice, users need to understand not only whether a run succeeds or fails, but also where repeated executions diverge in their coordination strategies, how those divergences emerge through task planning and execution, and how they can be traced back to concrete evidence in the execution logs. However, current debugging workflows still rely heavily on inspecting individual execution logs in isolation, offering limited support for cross-run alignment.

To establish a solid foundation for designing \tool, we conducted a formative study with five developers (E1--E5) who had hands-on experience in building, evaluating, and debugging LLM-based multi-agent systems (\autoref{tab:participants}).
E1 was a Ph.D. student focusing on healthcare-oriented multi-agent systems, including patient monitoring and clinical decision support, with experience in 3 projects.
E2, a master’s student and engineer, specialized in industrial AI-agent systems, working on manufacturing process optimization and collaborative robotics, with experience in 4 projects.  
E3, a Ph.D. student, conducted research on multi-agent systems across multiple domains, such as supply chain simulations and intelligent logistics, with experience in 5 projects.  
E4, an industry professional, focused on AI coding evaluation and multi-agent assessment workflows, including automated code testing and agent performance benchmarking, with experience in 6 projects.  
E5, also an industry professional, was deeply involved in research-oriented multi-agent projects, such as agent collaboration strategies and coordination in experimental settings, with experience in 5 projects.
We held regular biweekly meetings with these participants over the past six months to elicit design requirements for our system and iteratively refine it. This study was approved by the IRB. Based on the interviews, we distilled five analytical tasks as follows.

\begin{table}[h]
\centering
\caption{The demographics of participants in our formative study.}
\begin{tabular}{l l c l c}
\toprule
ID & Gender & Age & Specialization & \#Projects \\
\midrule
E1 & Female & 24 & Research & 3 \\
E2 & Male   & 21 & Industry & 4 \\
E3 & Male   & 23 & Research & 5 \\
E4 & Female   & 27 & Industry & 6 \\
E5 & Female & 25 & Industry & 5 \\
\bottomrule
\end{tabular}
\label{tab:participants}
\end{table}

\textbf{T1 Provide a high-level summary of outcomes across multiple runs.}
Participants consistently expressed the need for an overview that helps them quickly compare repeated runs before diving into detailed logs. For example, E4 emphasized the importance of first checking whether each run completed as expected and whether the overall execution remained stable across repetitions. E5 similarly wanted a concise way to compare successful and failed runs before investigating specific stages in more detail. These observations align with prior work on multi-agent system debugging, which highlights the importance of high-level summaries for navigating complex execution traces \cite{10.1145/3706598.3713581}. These findings suggest that the system should begin with a high-level summary of outcomes across multiple runs, enabling experts to identify suspicious runs or stages for further inspection.

\textbf{T2 Extract and identify common information nodes across runs.}
Participants emphasized the need to decompose complex tasks into smaller, comparable units across runs. E1 explicitly asked for a checkpoint mechanism that breaks a task into smaller nodes so that each stage can be checked against expectations. 
E4 similarly asked for comparisons centered on key nodes, enabling quick identification of where path deviations occur. This desire for structured task decomposition mirrors approaches in event sequence analysis, where breaking down complex processes into aligned stages enables systematic comparison ~\cite{monroe2013eventflow,magallanes2022Sequen-C}. These findings suggest that the system should identify common checkpoints or key nodes across runs as a basis for comparison.

\textbf{T3 Support cross-run comparison of node-level outcomes and differences.}
Once checkpoints (i.e., information nodes) are identified, participants emphasized the importance of systematically comparing node-level outcomes across multiple runs to gain deeper insights into the system's behavior. For example, E3 highlighted the need to examine differences in outputs, assess the distribution of these differences across repeated executions, and identify execution paths that are abnormal or rarely observed. Similarly, E4 stressed that clearly marking abnormalities and annotating the severity of errors at key nodes would greatly aid in quickly detecting critical issues and prioritizing investigation.  
These observations suggest that an effective system should not only support direct comparison of node-level outcomes across runs but also provide mechanisms for visualizing differences, distributions, and potential anomalies, thereby enabling users to better understand, diagnose, and reason about the behavior of complex multi-agent processes.

\textbf{T4: Enable analysis of agent actions leading to an information node.}  
Participants emphasized that merely identifying divergent nodes was insufficient; a comprehensive understanding requires analyzing the execution context that produced each outcome.
For instance, E2 noted that different runs might employ distinct sequences of actions to reach the same information node, and that comparing these behaviors is essential for identifying the root causes of errors.
E5 further emphasized the importance of systematically recording task assignments and execution sequences across runs. 
These insights suggest that the system should provide mechanisms to trace and analyze the sequence of actions leading to a selected information node. Such capabilities would enable users to systematically compare different runs, assess whether divergences are associated with retries, contextual failures, or strategy modifications, and ultimately facilitate rigorous diagnosis of complex multi-agent behaviors.

\textbf{T5: Present raw log data of agent actions.}  
While participants valued summaries and cross-run comparisons, they consistently emphasized that detailed diagnosis requires direct access to raw evidence. E3 requested an overview-to-detail workflow that allows users to zoom from node-level summaries into individual messages, agent-level inputs and outputs, and the original logs. E4 highlighted the need to inspect different types of log data, including the specific tasks completed by each agent, granular action records, and the resulting outcomes.
Similarly, E5 emphasized linking high-level summaries to detailed logs, particularly when errors occur. These observations suggest that the system should preserve access to raw agent log data while supporting progressive drill-down from overview to fine-grained detail.

\section{Information Node}

\begin{figure}[ht]
    \centering
    \includegraphics[width=\linewidth]{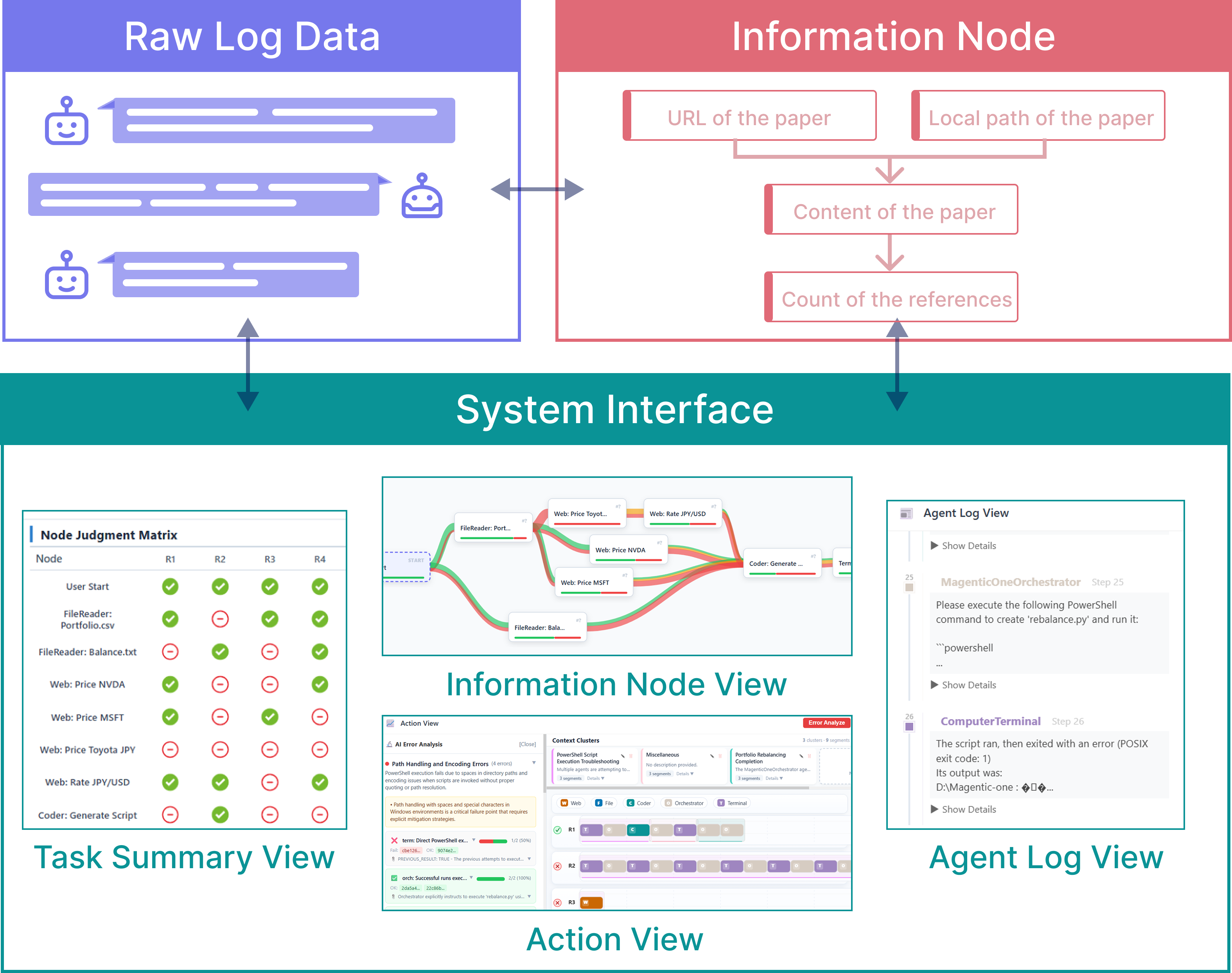}
    \caption{Conceptual framework of \tool illustrating the relationship between raw execution logs, information nodes, and the system interface. The upper left shows raw log data generated by agents. The upper right illustrates extracted information nodes, which represent a semantically meaningful task milestone abstracted from raw logs. 
    The lower part depicts the system interface, which operationalizes this abstraction through four coordinated views.}
    \label{fig:framework}
\end{figure}

To support debugging of behavioral inconsistency in LLM-based agentic systems across multiple runs, \tool introduces information nodes as an intermediate semantic abstraction between low-level execution logs and high-level task outcomes (\autoref{fig:framework}). 

This abstraction is inspired by findings from our formative study, in which developers consistently reasoned about inconsistency by first comparing the information acquired by different runs at the same task stage, before examining how that information was obtained.

Conceptually, an information node is a developer-facing analytical abstraction that represents a task-relevant informational milestone.
It corresponds to a point in execution where, from a debugging perspective, the system can be considered to have obtained enough intermediate information to move on to a different sub-goal.
Such a milestone is not defined by a specific action, but by a change in the system’s informational state.
Accordingly, instead of tracking low-level actions (e.g., \textit{``executing a Python script to parse a file''}), an information node captures the higher-level informational outcome achieved (e.g., \textit{``extracted the database schema''}), regardless of how that outcome was produced.
Specifically, our formative study revealed two key properties that developers consider essential when identifying appropriate information nodes during debugging:
\begin{itemize}
    \item \textbf{Recurrence and Abstraction.}
    Developers observed that different agent runs might follow divergent execution paths to solve the same task, and that extracting every piece of intermediate information leads to fragmented and noisy traces.
    They therefore emphasized the importance of focusing on \emph{recurrent} informational milestones—those that appear consistently across multiple runs.
    By centering analysis on such recurrent information nodes, they can abstract away incidental execution differences while preserving the core informational steps essential to task completion.
    This abstraction enables meaningful alignment and comparison of agent behaviors across runs.

    \item \textbf{Flexible Granularity.}
    Our formative study further indicated that developers do not reason about informational milestones at a fixed level of detail.
    Instead, they adjust the granularity of information nodes according to the debugging context.
    At early stages, developers often prefer coarse-grained nodes to obtain a high-level overview of task progress, while later stages of debugging often require refining and adjusting these nodes into finer-grained informational milestones.
\end{itemize}

However, although reasoning in terms of information nodes reflects developers’ natural debugging strategies, identifying and maintaining appropriate information nodes during the diagnosis process is non-trivial.
In practice, developers must manually scan long and heterogeneous execution traces, infer where meaningful informational milestones occur, and ensure that these milestones are consistently defined across multiple runs.
This process is time-consuming, error-prone, and difficult to scale as the number of runs or the complexity of agent behaviors increases.
To address this challenge, we introduce a recommendation pipeline that automatically proposes candidate information nodes from execution traces in centralized multi-agent systems.
These candidates are intended to reduce developers’ manual effort by surfacing plausible informational milestones that frequently occur across runs.
Developers can then select, refine, merge, or discard recommended nodes according to their specific debugging goals, such as obtaining a coarse overview of task progression or conducting a fine-grained inspection of a particular failure mode. 
In the following, we first describe the architecture of centralized multi-agent systems considered in this work.
We then explain how we leverage this structure to recommend information nodes from agent interactions and execution traces, and how these recommendations support flexible, human-in-the-loop debugging.

\subsection{Centralized Multi-agent Systems}


LLM-based multi-agent systems can be organized under different coordination architectures, depending on how planning, communication, and control are distributed among agents. 
Prior work commonly adopts one of four high-level architectures: hierarchical systems, centralized systems, decentralized systems, and systems with a shared message pool~\cite{aryal2026survey, Yao2025Survey}.
Among these designs, centralized multi-agent systems are one of the most widely used architectures in practice. 
In centralized systems, a central orchestrator decomposes the user request into intermediate goals, assigns subtasks to specialized worker agents, monitors execution progress, and decides whether to continue, revise strategies, or terminate. 
Therefore, the orchestrator maintains a global view of task progress; its messages provide a compact and semantically rich summary of the system’s intermediate states, including the very \textit{state transitions} that define our information nodes, as well as progress assessments and strategy revisions.

\subsection{Information Node Recommendation}

Information node construction in \tool follows a human-in-the-loop design.
The system automatically proposes candidate nodes from execution traces, while only developer-confirmed nodes will be treated as diagnostic abstractions in subsequent inconsistency analysis. The concrete extraction steps are as follows.

\textit{Step 1: Orchestrator message organization.}
In centralized multi-agent systems, the orchestrator is responsible for issuing high-level instructions (e.g., \textit{``find the URL corresponding to this article''}) that specify what information should be obtained and through which actions in the next step, while worker agents execute these instructions and return results.
As a result, orchestrator messages more directly reflect task intent, progress, and shifts in informational states, whereas worker-level traces often contain large amounts of low-level, intermediate, or noisy information.
To focus on task-relevant milestones, \tool therefore extracts orchestrator-centered traces as the primary input for information node construction.
This abstraction step filters out low-level execution noise and yields a compact sequence of coordination decisions that more clearly reflects task progression.

\textit{Step 2: Semantic change–driven segmentation.}
Given the abstracted orchestrator trace, \tool identifies the \textit{semantic state transitions} (i.e., the boundaries between informational milestones) by tracking semantic changes over time.
Each orchestrator message is encoded as an embedding vector, and semantic continuity between adjacent messages is measured along the trace.
Abrupt changes in this semantic trajectory, indicated by the cosine similarity between adjacent embeddings falling below a predefined threshold, are treated as candidate state transitions. This mathematically captures the moment the system acquires sufficient information to complete a subtask, shift to a new intermediate objective, or revise its strategy after failure.
Based on these transition points, each run is partitioned into a set of segments representing distinct informational states.
The resulting segments, each corresponding to a distinct informational state within a single run, are then used as input for candidate node extraction.

\textit{Step 3: Candidate information node extraction.} Considering that embedding-based segmentation can be sensitive to threshold choices and local semantic noise, the system prompts an LLM to generate natural language summaries from the original textual data associated with each cluster, thereby bridging the raw text and the operational definition of an information node.
These summaries serve as candidate descriptions, not final nodes. To reduce redundancy and improve usability, the system then performs a lightweight consolidation step by grouping semantically similar summaries together. This consolidation is used solely to merge obviously overlapping candidates and present developers with a manageable set of proposals.
Importantly, \tool does not attempt to infer a strict hierarchy or ontology of informational states. Ambiguities and borderline cases are intentionally preserved for human judgment rather than resolved automatically.

\textit{Step 4: Node selection and refinement.}
Automatically generated information nodes are treated as suggestions rather than final outputs.
Developers can review and refine the proposed nodes by editing their descriptions, merging semantically similar nodes, splitting overly broad nodes, or adding and removing nodes as needed to better reflect task-specific informational milestones.
Only the resulting confirmed node set, after this manual review and refinement process, is used in subsequent analyses.
This human-in-the-loop confirmation step ensures that information nodes remain interpretable, task-relevant, and aligned with developers' understanding of the system behavior.

\section{InconLens}

\begin{figure*}[h]
    \centering
    \includegraphics[width=\linewidth]{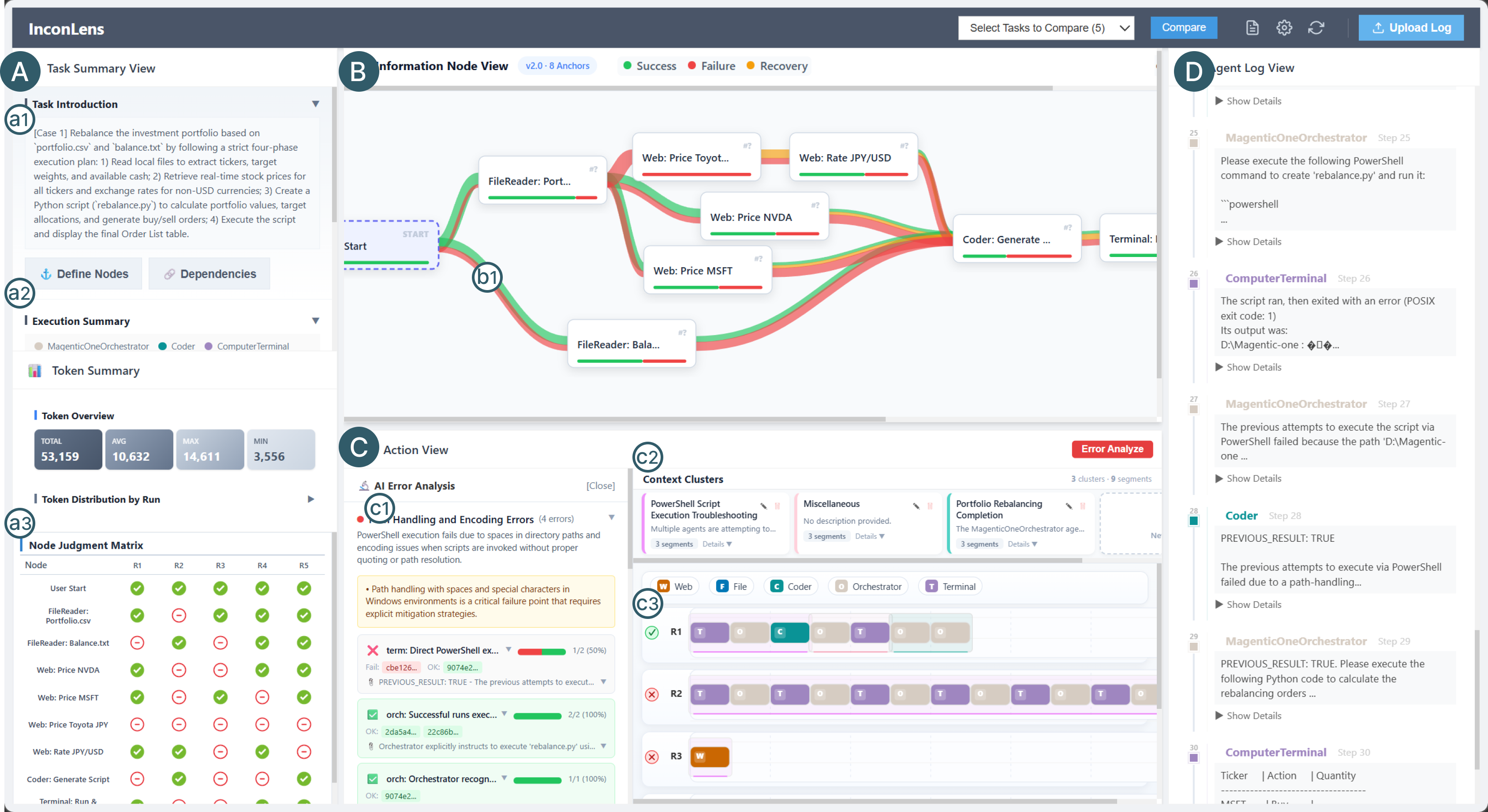}
      \caption{Overview of the \tool interface for diagnosing agent behavioral inconsistency across multiple runs. The system consists of four tightly coordinated views. (A) The Task Summary View provides a run-level overview of repeated executions. This view also supports the extraction, refinement, and dependency specification of information nodes.
  (B) The Information Node View visualizes the task as a sequence of information nodes connected by dependencies using a Sankey diagram, enabling users to identify divergent or failure-prone transitions.
  (C) The Action View supports detailed inspection of a selected node transition.
  (D) The Agent Log View presents the original execution logs in a chronological, step-by-step format.}
   \label{fig:teaser}
\end{figure*}

To support the diagnosis of agent behavioral inconsistency, \tool is organized into four coordinated views (\autoref{fig:teaser}). The \textit{Task Summary View} provides a run-level overview of multiple executions, including overall outcomes and a compact comparison of node-level judgments. It helps users quickly identify problematic runs or suspicious stages before reading detailed logs. The \textit{Information Node View} presents the extracted information nodes and their dependency structure, allowing users to understand the task’s informational flow and compare node-level outcomes across runs. Once a suspicious node is identified, the \textit{Action View} reveals the action sequences associated with that node, enabling users to compare how different runs approached the same milestone and where successful and failed trajectories began to diverge. Finally, the \textit{Agent Log View} provides direct access to the original execution records, allowing users to inspect raw evidence and verify the hypotheses formed in the higher-level views.
These four views form an overview-to-detail analysis pipeline. Users typically start with the run-level summary to understand overall patterns across multiple runs, then move to the information-node representation to identify the earliest meaningful divergence. They next inspect node-related action sequences to understand how the divergence emerged, and finally drill down into the raw logs for verification.

\subsection{Task Summary View}

The Task Summary View (\autoref{fig:teaser}-A) serves as the entry point of \tool and is designed for two complementary purposes: providing a concise run-level overview of repeated executions and supporting the construction of information nodes for downstream diagnosis \textbf{(T1)}. Rather than exposing detailed traces at the outset, this view presents high-level task context and compact execution summaries, while also allowing developers to extract, inspect, and refine the information nodes used in later views. To support these goals, the Task Summary View consists of three coordinated components: the task introduction component, the execution summary component, and the node extraction and dependency setup component.

\begin{figure}[h]
    \centering
    \includegraphics[width=\linewidth]{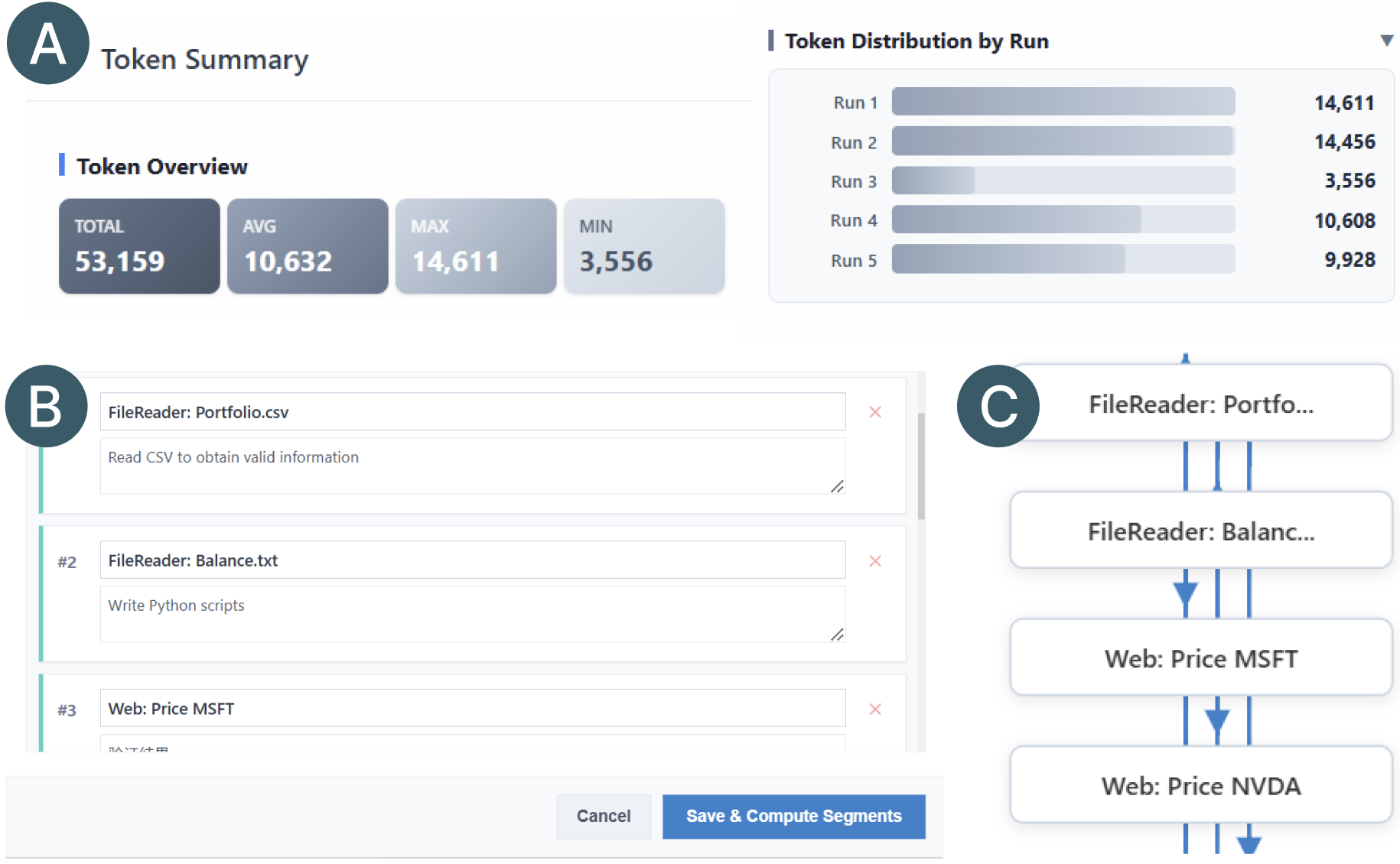}
    \caption{(A) Developers can examine a summary of token consumption. (B) Developers can generate and refine information nodes. (C) Developers can further inspect and modify the dependencies among the final determined information nodes.}
    \label{fig:nodeconstruction}
\end{figure}

The task introduction component provides users with a clear and immediate understanding of the current task the system is addressing (\autoref{fig:teaser}-a1), while the execution summary component offers a concise overview of each execution round, detailing the participating agents, token consumption, and other relevant operational information (\autoref{fig:teaser}-a2).
Furthermore, the final component allows users to define and select information nodes as starting points for subsequent analysis \textbf{(T2)}.
When users open the information node panel, the system automatically presents candidate information nodes extracted from execution traces.
This component supports lightweight refinement operations, including confirming, renaming, splitting, merging, generating, and refreshing nodes.
After developers determine the information nodes, the system visualizes the completion status of each information node across different runs (\autoref{fig:teaser}-a3 and \autoref{fig:nodeconstruction}-B).
Finally, our system can compute the dependency relationships among the determined information nodes through prompt engineering. Specifically, the system prompts an LLM with the task description and the finalized node set to propose prerequisite relationships between nodes, which are then presented for user review and manual correction.
To keep this process accurate, we also provide an interactive interface that allows developers to customize these dependencies (\autoref{fig:nodeconstruction}-C). Developers can directly edit node relations, specify task-specific dependencies, or import predefined information flows from external files. This design allows the dependency structure to reflect either the system-suggested information flow or users' prior knowledge of the task.

\subsection{Information Node View}

The Information Node View supports comparison of multiple runs at the level of task milestones. Specifically, it organizes the task into a set of information nodes and their relationships, enabling developers to examine how different runs progress through the same milestones and to identify recurring patterns, divergences, and failure or recovery behaviors across multiple runs \textbf{(T3)}.

\begin{figure}[h]
    \centering
    \includegraphics[width=1.0\linewidth]{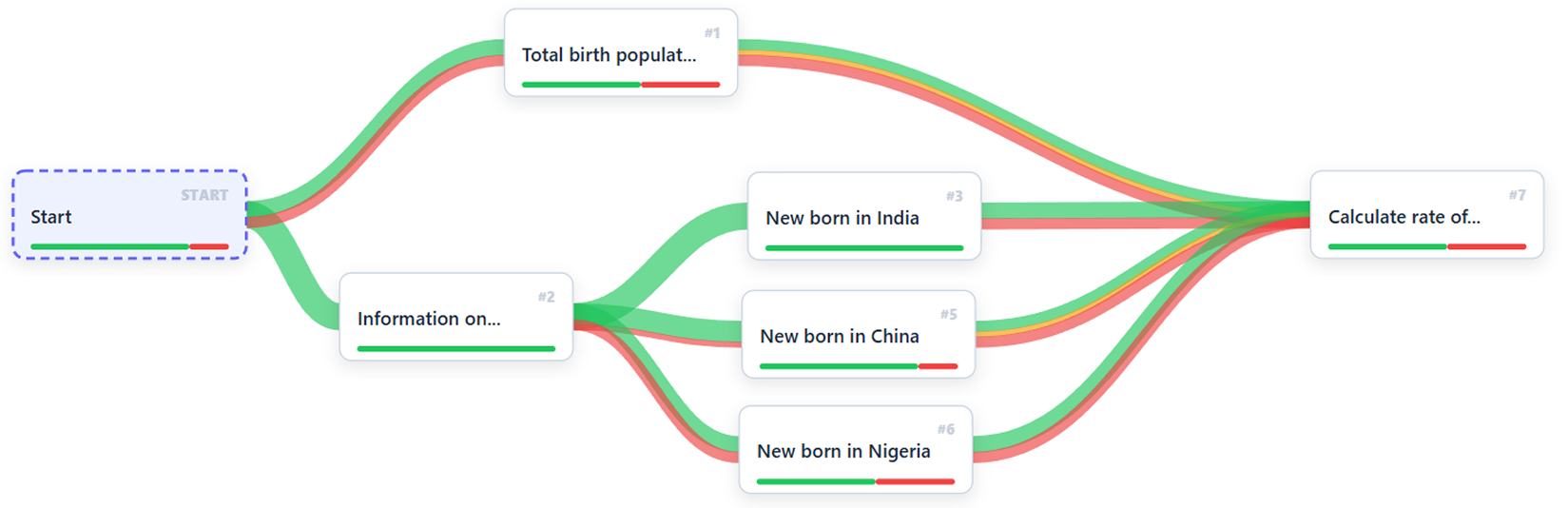}
    \caption{Sankey-based visualization of execution runs across information nodes, showing how tasks progress through milestones, the outcomes of transitions, and the relative frequency of different execution paths.}
    \label{fig:informationflow}
\end{figure}

The Information Node View uses a Sankey diagram to summarize how execution runs progress through different information nodes (\autoref{fig:informationflow}). The links between nodes represent transitions from one milestone to the next. The diagram is laid out from left to right, offering an intuitive overview of the overall task flow. Links are color-coded by outcome: green indicates successful transitions, red denotes failures that were not recovered, and orange represents transitions that initially failed but later succeeded (e.g., through retries or repairs). The thickness of each link reflects the number of runs following that path, allowing users to quickly distinguish stable execution paths from more problematic ones.
The diagram supports analysis at two complementary levels. At a global level, users can first grasp the overall task structure and identify the main execution branches. At a local level, they can zoom in on specific transitions to spot anomalies, such as links with a high failure rate or branches where many runs require recovery. By doing so, the design encourages users to move beyond inspecting individual runs and instead focus on recurring patterns across multiple runs at the same information node.

\begin{figure}[h]
    \centering
    \includegraphics[width=\linewidth]{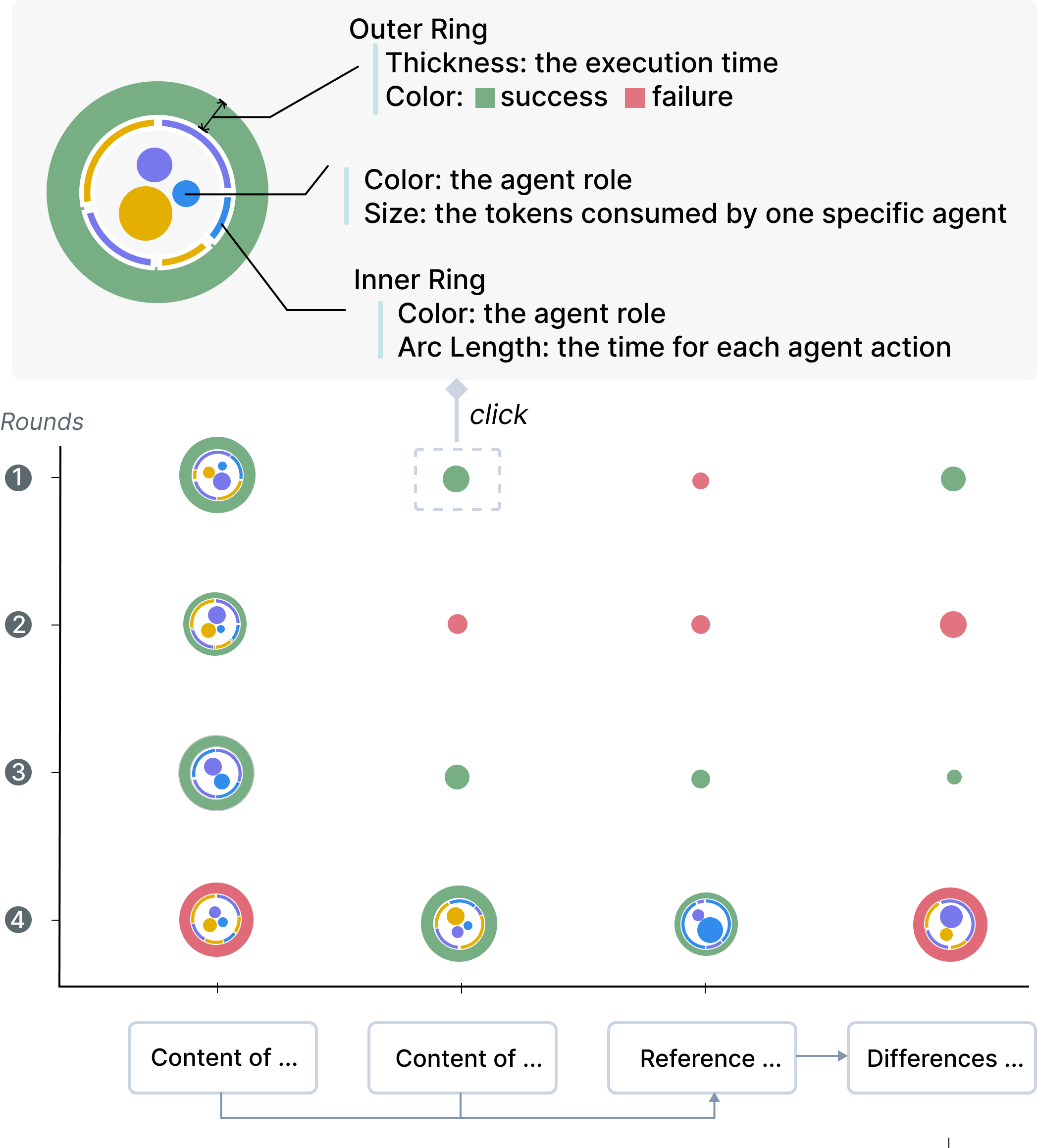}
    \caption{The initial glyph-based design summarizing run–node pairs. However, experts found the dense encoding overwhelming, motivating a flow-oriented approach that highlights key nodes first and defers detailed analysis to the Action View.}
    \label{fig:alternative}
\end{figure}

\textbf{Justification.}
Initially, we explored a glyph-based design (\autoref{fig:alternative}) in which each run–node pair was summarized by a compact circular mark encoding correctness, error magnitude, round count, agent execution order, and agent contribution. Although this design preserved rich node-level information in a small space, expert feedback indicated that the density of encoded attributes was cognitively overwhelming.
Through our observations, we found that developers typically follow a top-down debugging strategy: they first try to locate where problems occur before investigating why they occur. Based on this insight, we designed a flow that helps users quickly identify which information nodes are most likely to require further inspection, while deferring more fine-grained details to the Action View, where users can examine detailed agent actions and evidence on demand.

\subsection{Action View}
The Action View (\autoref{fig:teaser}-C) enables detailed inspection of how different runs traverse a selected transition in the Information Node View \textbf{(T4)}. When a user selects a Sankey link, this view surfaces the execution content associated with that transition and organizes it into three coordinated components: an \textit{AI Error Analysis} panel that summarizes recurrent failure patterns (\autoref{fig:teaser}-c1), a \textit{Context Clusters} panel that groups semantically similar execution segments (\autoref{fig:teaser}-c2), and an aligned sequence panel that visualizes how different runs progressed through the selected stage using actions from different agent types (\autoref{fig:teaser}-c3). Together, these components allow users to compare successful and failed actions of the same transition, identify repeated retries or detours, and form concrete hypotheses about why outcomes diverge before drilling down into raw logs for verification.

The first panel, \textit{AI Error Analysis}, provides a compact overview of major failure patterns observed within the selected transition (\autoref{fig:teaser}-c1). Rather than requiring users to manually inspect individual actions, this panel leverages LLMs to detect recurring error types, such as incomplete script generation, syntax errors, or inappropriate strategy selection. Each error type is linked to representative successful and failed examples, enabling users to quickly contrast ineffective behaviors with more effective ones and develop an initial explanation for why certain runs failed to complete the transition.

The second panel, \textit{Context Clusters}, groups action segments into recurring semantic contexts (\autoref{fig:teaser}-c2). Each cluster captures a common local theme within the selected transition, such as script generation, external information retrieval, or intermediate handling steps. These clusters provide a concise summary of the dominant behaviors that appear across runs, helping users assess whether the transition is characterized by a small number of consistent strategies or by multiple competing patterns. By abstracting over raw traces, this panel reduces complexity and facilitates comparison across runs.

The third panel presents aligned action sequences for multiple runs (\autoref{fig:teaser}-c3). Each row represents a single run, and each colored block corresponds to an action segment produced by a specific agent type (e.g., Web, File, Coder, Orchestrator, or Terminal). Read from left to right, these sequences reveal how runs progressed through the selected transition. Comparing rows allows users to quickly identify differences in action order, repeated attempts, missing steps, or prolonged detours. For instance, a successful run may exhibit a short, direct sequence, whereas a failed run may show repeated actions of the same type, indicating that the system became stuck in an ineffective strategy.

\subsection{Agent Log View}
The Agent Log View (\autoref{fig:teaser}-D) provides direct access to the original execution records and serves as the final verification layer of the diagnosis workflow \textbf{(T5)}. It presents agent log data in a structured, step-by-step chronological format, with each entry explicitly associated with a specific agent and execution step. This temporal organization allows users to reconstruct the complete execution process and understand how decisions and actions evolved over time.
Importantly, the view preserves the raw log content without abstraction or summarization, enabling users to inspect concrete details such as issued instructions, intermediate outputs, tool interactions, and execution errors. By grounding the analysis in original execution evidence, the Agent Log View supports careful validation of hypotheses generated from higher-level views and helps users confirm whether observed failures arise from agent reasoning, tool limitations, or external constraints.
\section{Case Study}

To demonstrate the effectiveness of our system in real-world diagnostic settings, we invited expert E4, who had previously participated in our design study, to conduct an in-depth analysis using our system. In this case study, E4 focused on diagnosing agent behavioral inconsistency in MagenticOne~\cite{fourney2024magentic}, which is a representative multi-agent system developed by Microsoft and widely used for coordinating complex, multi-step tasks.

Specifically, MagenticOne adopts a centralized, orchestrator-centric architecture, in which a single orchestrator agent is responsible for task decomposition, coordination, and progress control. Surrounding the orchestrator are several specialized LLM-based agents with distinct capabilities, including WebSurfer for retrieving external web information, FileSurfer for accessing and inspecting local files, Coder for generating executable scripts, and ComputerTerminal for executing commands and programs.
E4 selected an investment portfolio rebalancing case (\autoref{fig:teaser}-a1), a task that naturally requires coordinated reasoning across multiple agents, external information sources, and executable tools.
This task aims to rebalance an investment portfolio by determining which assets to buy or sell to reach a desired allocation, given the current holdings and available cash. 
Concretely, a typical correct execution path can be described in several steps. Starting from a record of the current portfolio and available cash stored in local files, the agents first gather the necessary external information, such as current market prices and currency exchange rates. Using this information, they then decide how the portfolio should be adjusted and translate this decision into a runnable program. Finally, the program is executed to produce a clear list of buy and sell orders, representing the final rebalancing outcome.

\begin{figure}[h]
    \centering
    \includegraphics[width=0.8\linewidth]{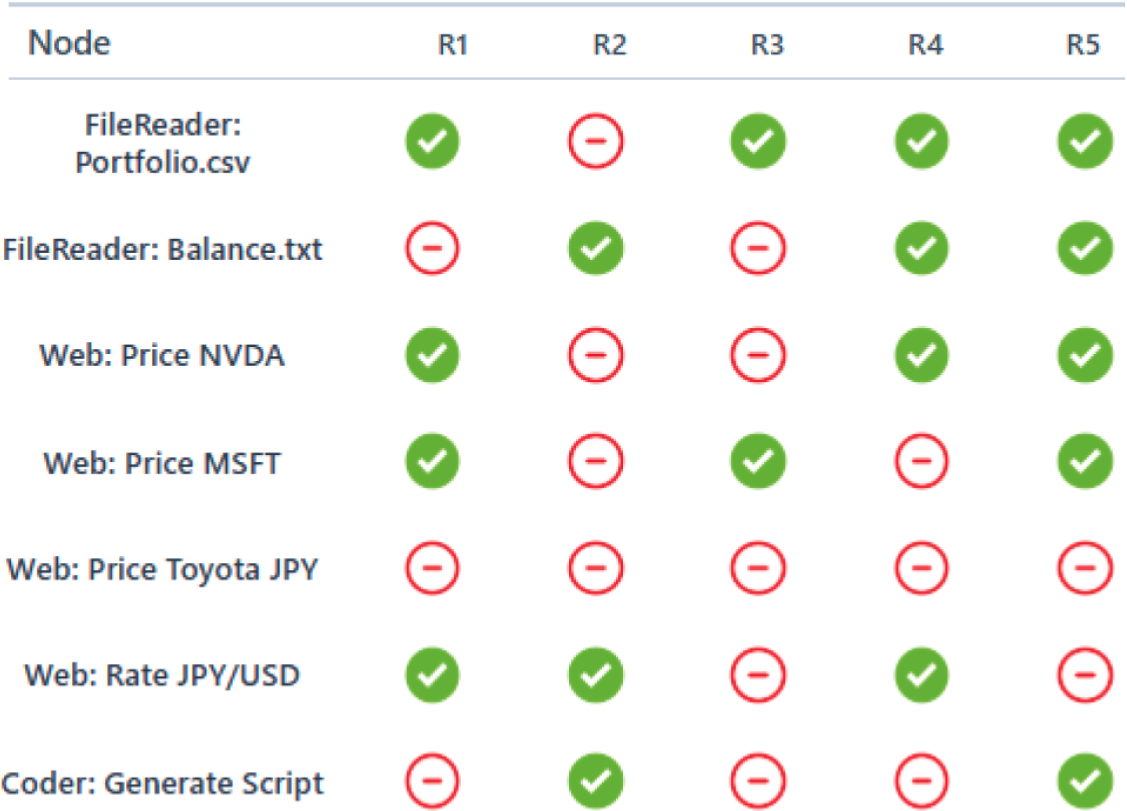}
    \caption{The determined information nodes were then evaluated based on whether they were completed in each run.}
    \label{fig:case1}
\end{figure}

\textbf{Construct information nodes for subsequent analysis (T2).}
After importing the raw logs from multiple runs, she first inspected the Task Summary View (T1) and observed that none of the runs had completed the overall task end-to-end. To understand whether these failures stemmed from a common cause, she conducted an initial comparison across runs. She then constructed a set of information nodes to capture task-relevant milestones, and iteratively refined them based on the task requirements and her domain knowledge. 
After computing information node completion across runs, she found that different runs exhibited varying completion patterns for specific information nodes (\autoref{fig:case1}), indicating that they failed at different stages of information acquisition. This suggested that the observed failures were not uniform, but rather arose from distinct error sources across runs.

\begin{figure}[h]
    \centering
    \includegraphics[width=1\linewidth]{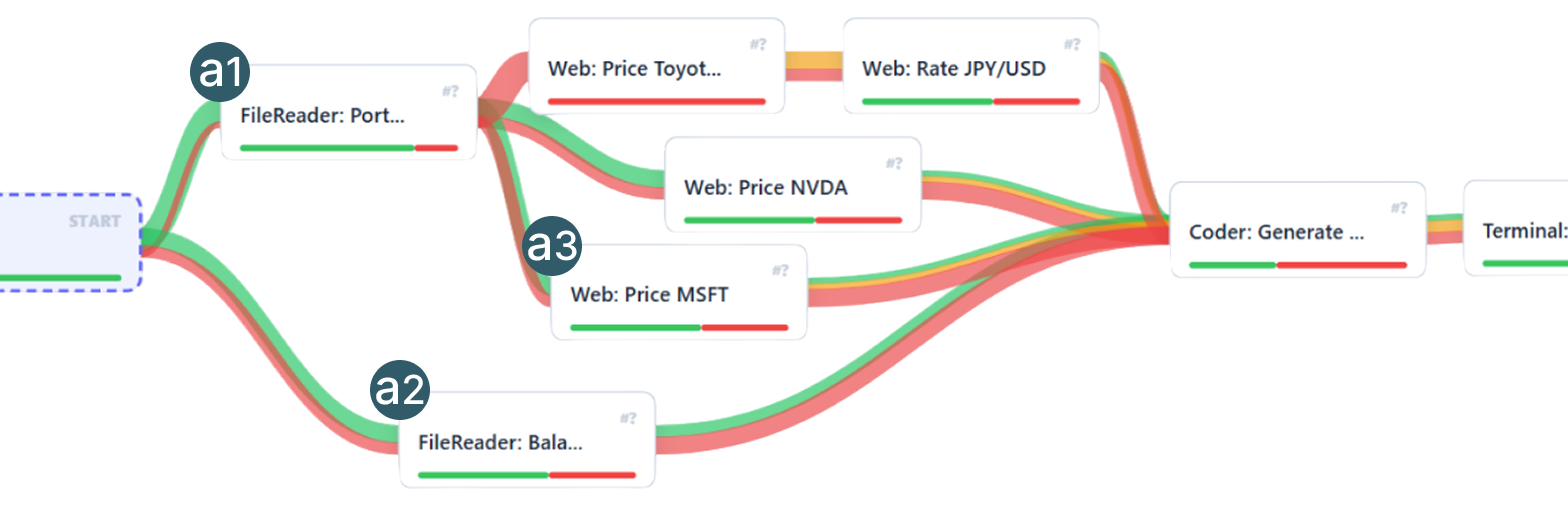}
    \caption{Node-level outcome overview for the portfolio rebalancing task across multiple runs in the Information Node View.}
    \label{fig:case2}
\end{figure}

\textbf{Explore the node-level outcomes (T3).}
E4 then proceeded to examine the node-level outcomes across multiple runs. By inspecting the node-level diagram (\autoref{fig:case2}), she was able to form a more structured understanding of how different information nodes depended on one another, as well as how consistently each node was completed across runs. This overview helped her move beyond individual execution traces and reason about systematic patterns of success and failure at the level of task-relevant information.
During this process, she first confirmed that the failures at nodes a1 and a2 were largely caused by the FileSurfer’s inability to correctly locate required file paths. However, rather than stopping at this localized issue, she became increasingly interested in the execution flow from a1 to a3. She observed that after successfully accessing the necessary files, different runs diverged in their subsequent behavior: some runs were able to continue and perform the required information search, while others stalled or failed at later steps.

\textbf{Inspect agent actions for failure identification (T4).} 
To understand why certain information nodes failed while others eventually recovered, E4 drilled down into the Action View for a more fine-grained inspection of agent behaviors. She first examined the AI-generated error summary, which synthesized recurrent failure patterns across runs. The summary highlighted that many agents repeatedly attempted to retrieve information directly from websites using the WebSurfer, becoming stuck in repetitive search loops without making meaningful progress. As a result, these runs failed during the information-gathering stage, even though no explicit execution errors were raised.
In contrast, the summary also pointed out that some partly successful runs exhibited a clear strategy shift. Rather than continuing direct web interactions, the orchestrator redirected the workflow toward a scripting-based approach, where agents generated code to programmatically fetch the required market data and then executed the script to complete the lookup. This change in strategy allowed those runs to bypass the limitations of interactive web browsing and successfully obtain the needed information.

\begin{figure}[h]
    \centering
    \includegraphics[width=\linewidth]{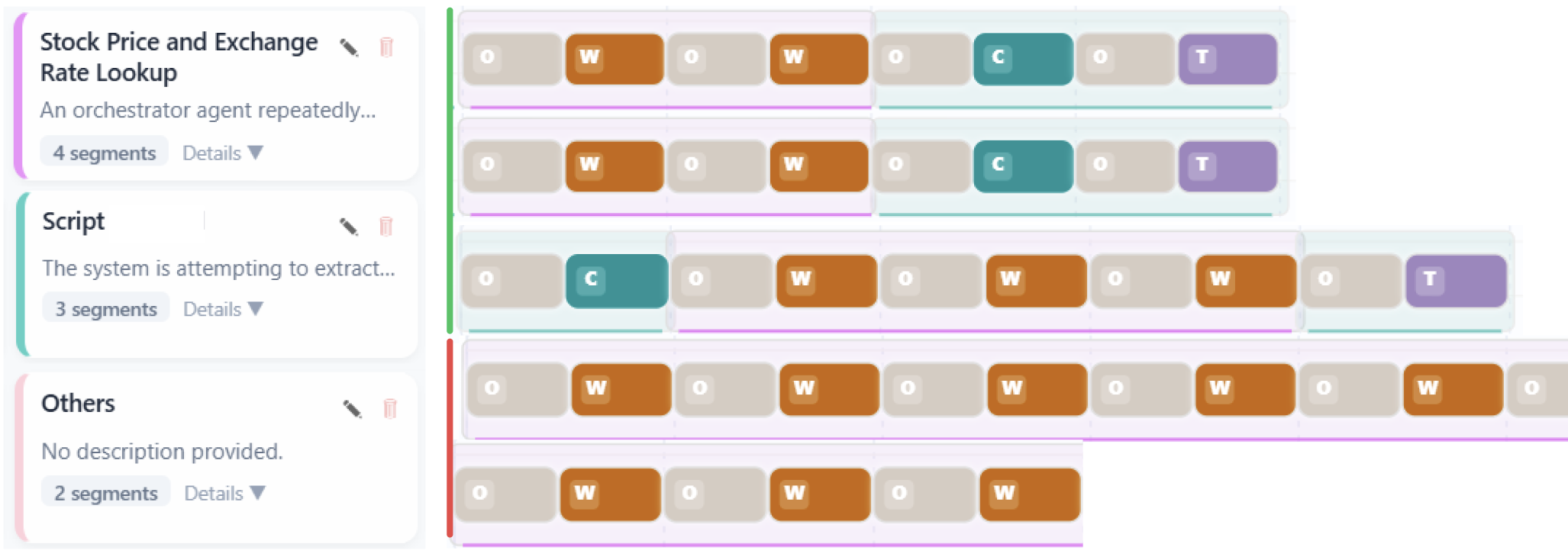}
    \caption{The left panel summarizes context clusters that group semantically similar action segments, while the right panel aligns action sequences from multiple runs along the same task stage.}
    \label{fig:case3}
\end{figure}

Motivated by this contrast, E4 proceeded to inspect the detailed action sequences underlying these outcomes (\autoref{fig:case3}). The system organized low-level actions into two prominent context clusters: one capturing sequences dominated by website-based searching through the WebSurfer, and another representing sequences in which agents generated and executed scripts to crawl or retrieve data programmatically. By comparing five representative execution paths across these clusters, she observed that failed runs tended to remain confined to the web-search cluster, repeatedly issuing similar actions without progressing toward task completion. In contrast, relatively more successful runs at this stage transitioned from web exploration to script-based data retrieval, leading to more consistent completion of downstream information nodes.
To further investigate this behavior, E4 switched to the Agent Log View to inspect the agents’ low-level execution traces (T5). There, she observed that after repeated attempts by the WebSurfer to access the same websites, the agent was eventually blocked by reCAPTCHA challenges that required manual verification. Because the system lacked the ability to detect or resolve such human-in-the-loop barriers, these runs became stuck in repeated web access attempts without making progress.
This deeper inspection confirmed that the failures were not caused by missing information or incorrect task decomposition, but rather by the orchestrator’s inability to recognize unproductive interaction patterns and trigger an appropriate strategy shift in a timely manner. 
Based on this insight, E4 proposed that MagenticOne could be improved by incorporating explicit signals for stalled interactions (e.g., repeated web access attempts or human-verification barriers) and using these signals to trigger earlier strategy changes, such as transitioning from interactive web browsing to script-based data retrieval.

After identifying this failure, E4 continued to review other error patterns surfaced by the system, using the same workflow to quickly trace high-level failures back to concrete agent behaviors. This process demonstrated how the system supports an iterative diagnostic loop: enabling developers to move from aggregate outcomes, to task-relevant information breakdowns, and finally to fine-grained action- and log-level evidence. Such a structured analysis approach allows developers not only to pinpoint failure causes efficiently, but also to reason about targeted system improvements in a principled and systematic manner.

\section{Expert Interview}
To assess the effectiveness of \tool, we conducted a task-based expert evaluation with five participants (P1–P5), including four males and one female.
All of them had hands-on experience in building, evaluating, or debugging LLM-based agentic systems.
Specifically, the five participants included a range of expertise in LLM -based agentic systems and related fields. P1 is a software engineer with experience in five projects developing LLM-based multi-agent systems. P2 is a researcher who has contributed to seven projects in agentic system evaluation. P3 is a Ph.D. student with three projects focused on healthcare-related agentic systems. P4 is a postdoctoral researcher with four projects in autonomous agent design and multi-agent coordination. P5 is a software engineer with eight projects in multi-agent system implementation and debugging.
None of the participants had been involved in the system’s design or development. 
The primary goal of this study was to examine whether \tool can support the diagnostic reasoning workflow: forming a high-level understanding of system behavior, identifying divergences across repeated runs, tracing these divergences to specific agent actions, and verifying findings against raw execution logs.

\begin{table}[h]
\centering
\caption{The task list in our expert interview.}
\renewcommand{\arraystretch}{1.3}
\begin{tabular}{p{3cm}|p{7.2cm}}
\hline
\multicolumn{2}{c}{\textbf{Tasks (In order)}}                                                                                                                          \\ \hline
\multicolumn{1}{l|}{Task1} & Use the Task Summary View to form a high-level understanding of outcomes across multiple runs.                                                                 \\ \hline
\multicolumn{1}{l|}{Task2} & Generate, refine, and determine information nodes for subsequent analysis in the Task Summary View.                                \\ \hline
\multicolumn{1}{l|}{Task3} &  Use the Information Node View to understand the completion status of the task’s informational milestones and compare node-level outcomes across runs to identify where behavioral inconsistencies occur.               
    \\ \hline
\multicolumn{1}{l|}{Task4} & Analyze the agent actions that led to a problematic node’s success or failure across different runs in the Action View. \\ \hline
\multicolumn{1}{l|}{Task5} & Drill down into raw agent logs to verify a specific failure hypothesis in the Agent Log View.                \\ \hline
\end{tabular}
\label{tasklist}
\end{table}

Each evaluation session was conducted either online via Zoom with screen sharing or in person and lasted approximately 70 minutes. Sessions followed a structured flow designed to immerse participants in the system while capturing both quantitative and qualitative data. 
Participants first received an introduction to our project context, including the challenges of behavioral inconsistencies in LLM-based agentic systems.
Then, we introduced the four coordinated views of our system and interactions among them. 
They then freely explored the system to familiarize themselves with its interface and visualizations. 
This was followed by an introduction to the architecture of MagenticOne~\cite{fourney2024magentic}, a centralized multi-agent system developed by Microsoft that would be the target of debugging in the subsequent tasks. 
After that, participants were asked to complete five tasks (\autoref{tasklist}) through exploring a pre-loaded case from our case study: an investment portfolio rebalancing task executed by MagenticOne.
Following task completion, participants filled out post-task questionnaires, including the NASA-TLX~\cite{hart2006nasa}, to assess workload and perceived usability. Finally, they participated in semi-structured interviews, providing feedback on the usefulness and effectiveness of our system, and potential improvements or feature additions.
All participants (P1–P5) received approximately \$12 as compensation for their time after completing the study.

\textbf{System Effectiveness.}
Participants consistently emphasized that \tool significantly improved their ability to diagnose repeated-run executions, particularly when dealing with complex workflows and long execution traces. P1 remarked, \textit{``Before using this system, I would spend hours manually comparing logs line by line. Now, I can immediately see which runs failed and focus directly on the nodes that might be causing the problem. It saves a huge amount of time and mental effort.''} P2 highlighted the value of information nodes as an intermediate abstraction, stating, \textit{``These nodes act like checkpoints that are shared across runs. They let me reason about dependencies and recovery systematically, rather than jumping around in the raw data.''} P4 also mentioned that the system helped them effectively form initial hypotheses about divergence points before diving into the raw logs: \textit{``It’s like having a guided map. I can see where something might have gone wrong and only then check the logs to confirm my ideas.''} 
Collectively, these observations suggest that \tool effectively structures expert reasoning, thereby supporting more systematic and accurate diagnosis of behavioral inconsistencies in multi-agent LLM systems.

\textbf{Workflow, Interaction, and Visualization Support.}
Experts praised the overview-to-detail workflow as a core strength of \tool. P4 described the process in detail: \textit{``This system guided me step by step, structuring my previous debugging approach in a more systematic way. This stepwise process helps me gradually refine my understanding.''} 
P1 added that the system promotes a more structured diagnostic reasoning process.
P5 emphasized the system’s interactive features, noting, \textit{``Being able to click on nodes and see detailed information, along with clear labels and color cues, makes it much easier to track issues.''}
In addition, P1 and P3 highlighted that they particularly appreciated the flow diagram in the Information Node View, noting that it made the task’s information flow very clear. As P3 explained: \textit{``The flow diagram clearly shows how each piece of information depends on others, which makes it much easier to understand the overall task structure and identify where inconsistencies occur.''}
Overall, participants found that \tool effectively supports expert reasoning by providing clear visual abstractions, structured workflows, and intuitive cross-run comparisons.

\begin{figure}[h]
    \centering
    \includegraphics[width=\linewidth]{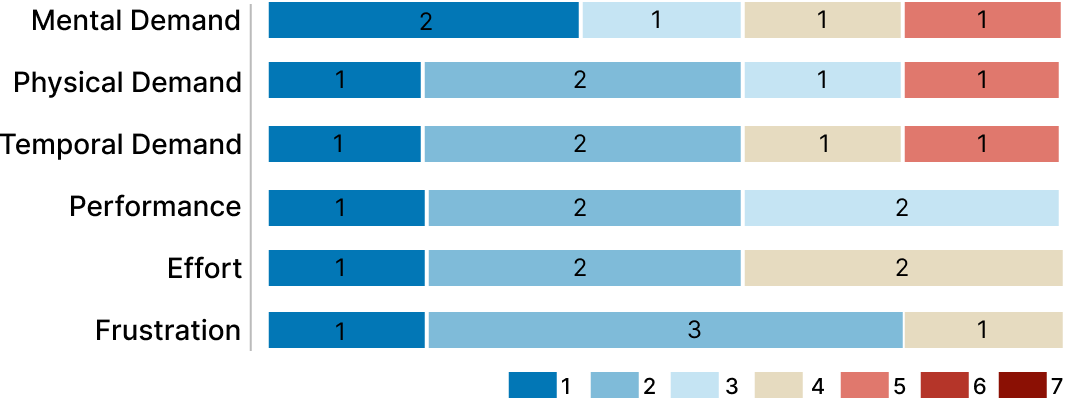}
    \caption{ The cognitive load results.}
    \label{fig:cognitive load}
\end{figure}

\textbf{Cognitive Load.}
Participants reported a generally manageable workload when using \tool to complete our tasks for the diagnosis of LLM-based multi-agent systems. 
NASA-TLX ratings (on a 1–7 scale) indicated low scores across most dimensions (\autoref{fig:cognitive load}): Mental Demand ($M = 2.8, SD = 1.6$), Temporal Demand ($M = 2.8, SD = 1.5$), Physical Demand ($M = 2.6, SD = 1.4$), Effort ($M = 2.6, SD = 1.2$), Frustration ($M = 2.2, SD = 1.0$), and Performance ($M = 2.2, SD = 0.7$). 
This suggests that participants found \tool easy to use and not overly taxing, indicating that the system imposes a low cognitive load while supporting the diagnostic tasks.
Moreover, participants (P1, P2, and P4) noted that the structured workflow—from high-level task summaries down to fine-grained action logs—reduced the need for constant context switching, a common source of mental strain when manually comparing execution traces. Our participants also remarked that the visual aggregation of failure patterns and the semantic clustering of action segments allowed them to maintain a clear mental model of the task progression without being overwhelmed by low-level details. The consistency of these subjective reports across participants further supports the interpretation that the abstraction layers introduced by \tool meaningfully reduce cognitive overhead.

\textbf{Suggestions.}
Participants also offered several valuable suggestions to further enhance \tool’s usability and effectiveness. For example, P2 and P4 mentioned that providing customizable color schemes or visual encodings for node states could make patterns across runs easier to perceive. P3 suggested adding optional summary tooltips for nodes, allowing users to quickly recall what each node represents without opening full details. P5 highlighted that supporting minor layout adjustments, such as collapsing or resizing panels, could help tailor the interface to different tasks or screen sizes.
\section{Discussion}
In this section, we will discuss the design implications, generalizability, and limitations of our work.

\subsection{Design Implications}
We elaborate on three design implications that emerge from our design exploration in diagnosing LLM-based agentic systems. 

\textbf{Constructing semantic abstraction layers for diagnosing LLM-based agentic systems.}
To alleviate the cognitive burden of navigating verbose and unstructured log data in LLM-based agentic systems, our design moves beyond raw, low-level text by introducing ``information nodes'' as an intermediate abstraction layer. This layer transforms fragmented execution traces into a sequence of high-level task milestones.
This way, granular actions can be aggregated semantically, enhancing users' understanding of those actions. 
Furthermore, by mapping diverse execution paths onto a unified semantic space, multiple runs can also be better aligned. Therefore, this abstraction enables developers to rapidly discern primary execution patterns and pinpoint behavioral inconsistencies that would otherwise be obscured in raw logs, thereby enhancing diagnostic precision and efficiency.

\textbf{Combining LLM automation with developer expertise to construct information nodes.}
Given the inherent complexity of agent behavior, fully automated generation of information nodes is often insufficiently accurate. Developers may also hold their own interpretations of tasks that are not captured by automated processes. Our system addresses this by enabling collaborative creation of information nodes between LLMs and developers. In this workflow, the LLM primarily proposes candidate nodes, while developers refine, modify, and validate these suggestions. The design explicitly supports developer intervention on top of automated recommendations. Specifically, it allows adjustments such as merging overlapping steps, splitting critical decision points, or relabeling nodes to reflect task-specific insights. This human-in-the-loop approach ensures that the resulting information nodes serve as a reliable foundation for subsequent analysis.

\textbf{LLM-assisted error summarization with human validation.} Our system leverages the generative capabilities of LLMs to automatically summarize potential errors and present them in the Action View.
This shifts humans from exploratory roles to validators of agent behavior. In prior work, such as DiLLS~\cite{sheng2026dills}, developers primarily relied on visualization-based hints to locate failures. In contrast, our approach delegates the initial summarization to LLMs. 
While LLM-generated summaries may occasionally contain hallucinations, our design and interaction mechanisms ensure that users can quickly trace back to the original log data, verifying the accuracy of the summaries and mitigating the impact of potential errors. This combination of automated synthesis and human validation has the potential to enhance both efficiency and reliability in failure analysis.

\subsection{Generalizability}
While \tool is currently designed for centralized LLM-based multi-agent systems, its underlying concept of ``information nodes'' offers a foundation that could be adapted to other agentic paradigms through specific enhancements.
However, the methodology for recommending and generating information nodes could be further adapted to accommodate diverse agentic architectural paradigms.
Specifically, for decentralized or peer-to-peer architectures where a global coordinator is absent, our methodology could be extended by incorporating a standardized self-reporting layer via prompt engineering. By guiding individual agents to output their perceived information and task intent in a structured format, the system could potentially synthesize these distributed insights into a unified execution trace for node extraction. In hierarchical systems, the current framework could be refined into a recursive diagnostic model, extracting nodes at different granularities to capture both high-level delegation and low-level execution. Furthermore, in shared message pool environments, the system could be adapted to treat the communication stream as a ``virtual orchestrator''. By applying time-windowed semantic clustering to asynchronous messages, it becomes possible to reconstruct the logical progression of tasks into aligned milestones.
Ultimately, these potential extensions suggest that the principle of using semantically aligned information nodes to diagnose behavioral inconsistencies can serve as a versatile starting point, capable of evolving alongside increasingly complex and diverse agentic infrastructures.

\subsection{Limitations and Future Work}
We identify two main limitations of our diagnosis system: the risk of hallucinations in LLM reasoning and difficulties in handling highly divergent execution paths.

\textbf{Potential hallucinations in LLM-based reasoning for information node evaluation.}
Our system relies heavily on the reasoning capabilities of LLMs to determine the completion status of information nodes, which inevitably introduces a risk of hallucinations. Our current mechanism mitigates this risk by providing comprehensive context information, such as previous information node completion states, relationship dependencies, and task descriptions. Interestingly, we observed a ``self-correction'' phenomenon. Since the diagnostic process involves multiple sequential judgments, even if a node’s completion status is initially misclassified, subsequent node evaluations might correct the earlier error based on updated execution traces, thereby reducing the impact of single-instance hallucinations. Nevertheless, this robustness may be challenged when processing highly ambiguous or logically overlapping logs, and fully eliminating hallucination-induced inaccuracies remains an open problem.
To mitigate this, multiple detection passes with a voting mechanism could be employed to reinforce robust node identification and reduce the impact of spurious signals. 

\textbf{Challenges in universal information node extraction for highly divergent execution pathways.}
Our work primarily targets tasks with relatively well-defined completion paths. However, in some open-domain tasks, agent execution paths can exhibit divergence across runs. Under such conditions, extracting a universally representative set of information nodes that accurately captures all runs becomes highly challenging. Nodes that are too coarse may miss critical execution details, while overly fine-grained nodes may lack cross-run commonality, resulting in fragmented views. In these scenarios, it becomes essential to engage more closely with developers to understand how they diagnose inconsistencies in agent behavior, including the strategies and heuristics they rely on. Insights from such human-in-the-loop investigations could inform the design of adaptive node extraction methods that better accommodate divergent execution pathways.
\section{Conclusion}
In this work, we presented \tool, an interactive visual analytics system designed to diagnose behavioral inconsistencies in LLM-based multi-agent systems. Recognizing that traditional debugging tools fall short in supporting cross-run comparison and semantic alignment of agent behaviors, we introduced the concept of information nodes as a high-level semantic abstraction. This abstraction enables developers to move beyond unstructured, verbose execution logs and instead reason about task progress in terms of recurring informational milestones that are shared across multiple runs. By aligning diverse execution trajectories onto a unified set of milestones, \tool facilitates systematic comparison of agent behaviors, helping users pinpoint where and why divergences occur.
Our evaluation demonstrates that \tool can facilitate the identification of behavioral drifts that are often obscured in traditional debugging environments.
Future research will focus on enhancing the robustness of node extraction by investigating more versatile extraction strategies. 
Ultimately, we believe that providing developers with structured, semantically-aware diagnostic tools is a critical step toward building more reliable, transparent, and steerable LLM-based autonomous systems.

\bibliographystyle{abbrv-doi-hyperref}
\bibliography{main}

\end{document}